\documentclass[%
 preprint,
 superscriptaddress,
 amsmath,amssymb,
 aps,
]{revtex4-2}

\usepackage{graphicx}
\usepackage{dcolumn}
\usepackage{bm}

\usepackage{siunitx}
\usepackage{braket}
\usepackage{geometry}
\usepackage{appendix}

\begin{document}

\title{Quantifying NV-center Spectral Diffusion by Symmetry}

\author{B. A. McCullian}
 \affiliation{School of Applied and Engineering Physics, Cornell University, Ithaca, NY 14853, USA}
\author{H. F. H. Cheung}
 \affiliation{Department of Physics, Cornell University, Ithaca, NY 14853, USA}
\author{H. Y. Chen}
 \affiliation{Department of Physics, Cornell University, Ithaca, NY 14853, USA}
\author{G. D. Fuchs}
\email{gdf9@cornell.edu}
 \affiliation{School of Applied and Engineering Physics, Cornell University, Ithaca, NY 14853, USA}
 \affiliation{Kavli Institute at Cornell for Nanoscale Science, Ithaca, NY 14853, USA}

\date{\today}

\begin{abstract}
The spectrally narrow, spin-dependent optical transitions of nitrogen vacancy (NV) center defects in diamond can be harnessed for quantum networking applications. Key to such networking schemes is the generation of indistinguishable photons. Two challenges limit scalability in such systems: defect-to-defect variations of the optical transition frequencies caused by local strain variation, and spectral diffusion of the optical frequencies on repeated measurement caused by photoexcitation of nearby charge traps. In this experimental study we undertake a group theoretic approach to quantifying spectral diffusion and strain, decomposing each into components corresponding to Jahn-Teller symmetries of the NV center. We investigate correlations between the components of strain, spectral diffusion, and depth from surface, finding that strain and spectral diffusion are each dominated by longitudinal perturbations. We also find a weak negative correlation between transverse static strain and total spectral diffusion suggesting that transverse strain provides some degree of protection from spectral diffusion. Additionally, we find that spectral diffusion becomes more pronounced with increasing depth in the diamond bulk. Our symmetry-decomposed technique for quantifying spectral diffusion can be valuable for understanding how a given nanoscale charge trap environment influences spectral diffusion and for developing strategies of mitigation.
\end{abstract}

\maketitle

\section{Introduction}

Quantum networking schemes rely on generation and distribution of entanglement via optical photons \cite{Childress2005, Childress2006, Reiserer2015, Faraon2011, Gao2015}. The spectrally narrow, spin-dependent optical transitions of nitrogen vacancy (NV) centers in diamond have enabled them to become a leading platform for such applications. Significant advances, including entanglement of NV centers with photons \cite{Togan2010}, quantum interference of photons from distinct NV centers \cite{Sipahigil2012, Bernien2012}, and heralded entanglement of remote NV centers \cite{Bernien2013} have enabled landmark loophole-free tests of Bell's inequality \cite{Hensen2015, Hensen2016}, demonstration of a three node quantum network \cite{Pompili2021}, and teleportation of a state across such a network \cite{Hermans_2022} using NV centers. However, two challenges limit scalability for such quantum networking schemes that are reliant on frequency matched photon generation from distinct NV centers. First, coupling to the local strain environment results in defect-to-defect variation of NV-center optical transition frequencies, necessitating a means of tuning the NV-center optical transitions \cite{Tamarat2006, Tamarat2008, Bassett2011a, Sipahigil2012, Bernien2012, Acosta2012, Schmidgall2018}. Second, photoinduced changes to the local charge trap environment on repeated measurement cause spectral diffusion of the NV-center optical transitions \cite{Jelezko2002, Tamarat2006, Shen2008, Fu2009, Ishikawa2012}. The need for pre- or post-selection schemes to mitigate spectral diffusion reduces the entanglement generation rate \cite{Robledo2011}.

Symmetry plays a key role in determining the optical transition frequencies of NV centers. Group theory has been used to identify the symmetry of the NV-center ground and excited states \cite{Maze2011b, Doherty2011a}, and perturbations of the NV center can be decomposed in terms of Jahn-Teller \cite{H.A.JahnandE.Teller1937} symmetries where perturbations of distinct symmetries cause different shifts of the optical transitions. Group-theoretic understanding of the NV center and its susceptibility to electric field fluctuations responsible for spectral diffusion has spurred the investigation of other defects in diamond with different symmetries that exhibit more robust optical transitions \cite{Iwasaki2017, Rose2018, Trusheim2019, Bradac2019, Bhaskar2020, Rugar2021}. Despite these advances, the NV center remains the leading defect-based platform for entanglement distribution \cite{Pompili2021}, and detailed characterization of the static strain and spectral diffusion of NV centers will impact defect engineering \cite{Kasperczyk2020, Yurgens2021, Chakravarthi2021} for generating entanglement.

We investigate the correlations between spectral diffusion and strain using symmetry resolved measurements of NV-center spectral diffusion. Previous work has identified that optical spectroscopy of the two excited-state orbital branches allows the quantification of strain along and transverse to the NV-center axis \cite{Grazioso2013, Lee2016, Chen2018}. We measure the spectral diffusion of optical transitions in both orbital branches and from these measurements are able to decompose spectral diffusion into components. We report measurements on 16 individual NV centers in bulk diamond. First, we quantify the components of static strain and spectral diffusion for each defect. Next, we use this set of measurements to calculate correlations between the different components of static strain and spectral diffusion, as well as the depth from the diamond surface. We uncover strong correlations suggesting that strain and spectral diffusion are both dominated by perturbations along the NV-center symmetry axis. We find weak correlation between transverse static strain and spectral diffusion. Additionally, we find that spectral diffusion becomes more pronounced deeper into the diamond bulk.

\section{Measuring Static Strain and Spectral Diffusion}

\begin{figure}[t]
\includegraphics[scale=0.9]{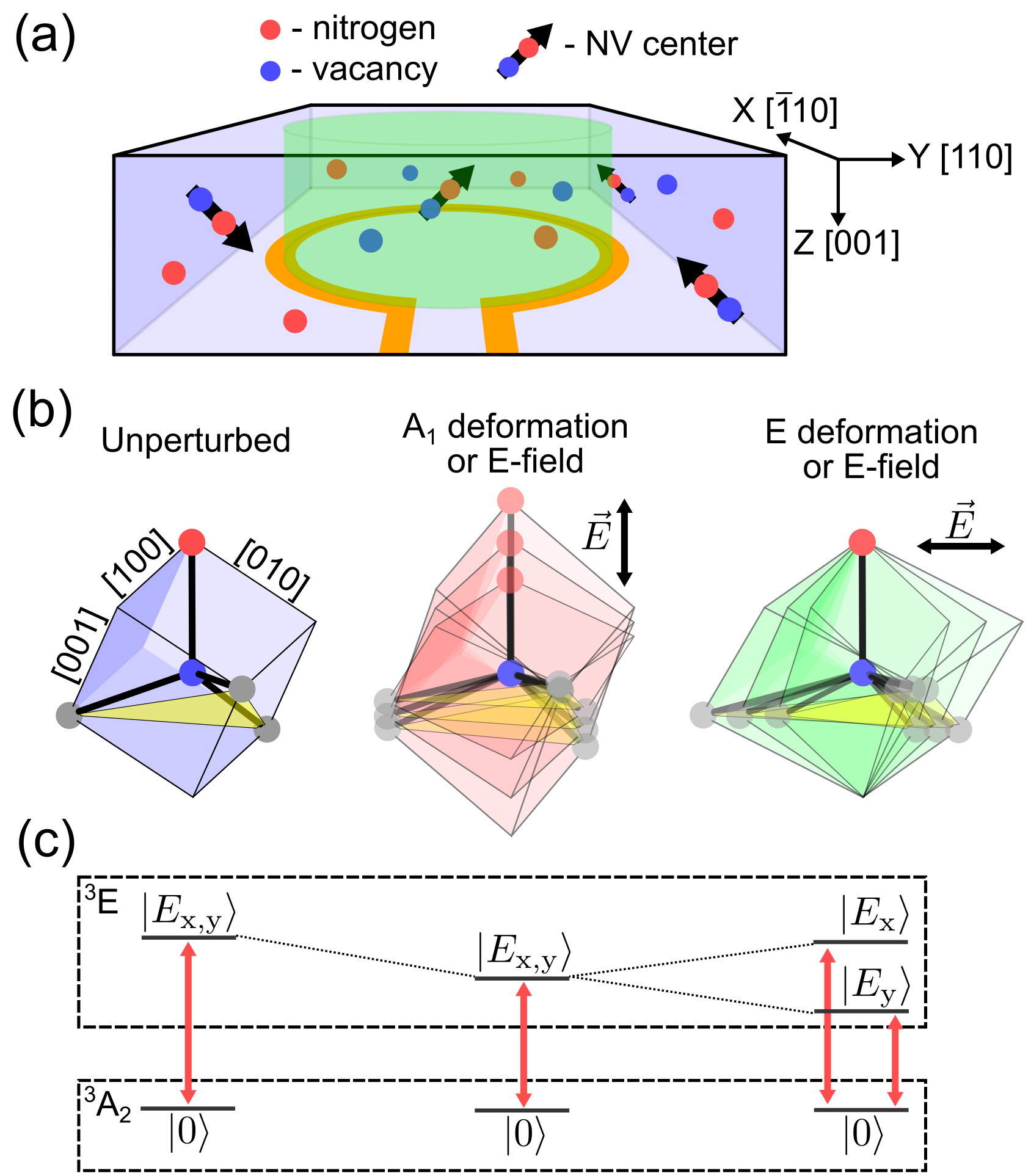}
\caption{\label{fig:Fig 1} (a) Schematic of diamond sample with spatially isolated single NV centers and nearby bulk charge traps. The green laser used for NV center spin and charge initialization can also couple to and modify the charge trap environment. A microwave loop antenna is used for NV-center electron spin resonance manipulation. (b) An unperturbed NV center has $C_{3v}$ symmetry and (c) a splitting between a ground $m_s = 0$ state and two-fold degenerate $m_s = 0$ excited states at 1.945 eV (637.2 nm) \cite{G.Davies1976, Jelezko2002}. $A_1$-symmetric (longitudinal) perturbation from strain or electric field globally shifts all ground to excited-state transition energies. $E$-symmetric (transverse) perturbation splits the excited-state orbital doublet into two branches. One example of $E$ type perturbation is illustrated.}
\end{figure}

Our sample [Figure \ref{fig:Fig 1}(a)] is a type IIa diamond with $\langle 100 \rangle$ oriented faces and $\langle 110 \rangle$-oriented edges purchased from Element Six. Individually addressable single NV centers were formed via electron irradiation and subsequent annealing \cite{MacQuarrie2015, Chen2018}. A microwave antenna (Ti 25 nm/Pt 225 nm) is patterned on the diamond for magnetic spin control. We perform photoluminescence excitation (PLE) spectroscopy measurements by tuning a red laser ($\sim$637.2 nm) across the resonant optical transition frequencies of the NV centers while counting photons emitted into the phonon sideband. All measurements are carried out in a helium flow cryostat at 10 K. We investigated 54 defects, 39 of which exhibited a PLE response. Issues with charge state stability of certain defects, as well as the mode hop free tuning range of our red laser ($\sim 20$ GHz) limited our study and analysis to 16 of the defects for which we were able to collect reliable signal.

The spin-preserving zero phonon line (ZPL) optical transitions of the negatively charged NV center ($\textrm{NV}^{-}$) are from orbital singlet, spin triplet ($^3\textrm{A}_{2}$) ground states \{$\ket{0}$,$\ket{\pm1}$\} to orbital doublet, spin triplet ($^3\textrm{E}$) excited states \{$\ket{E_1}$, $\ket{E_2}$, $\ket{E_y}$, $\ket{E_x}$, $\ket{A_1}$, $\ket{A_2}$\} at around 1.945 eV (637.2 nm) \cite{G.Davies1976, Jelezko2002}. We restrict our discussion to the two $m_s = 0$ excited states, $\ket{E_x}$ and $\ket{E_y}$, one of which is in either orbital branch. An unperturbed NV center has $C_{3v}$ point group symmetry [Figure \ref{fig:Fig 1}(b)] and degenerate $\ket{E_{x,y}}$ transitions \cite{Maze2011b, Doherty2011a}. Longitudinal perturbations from strain or electric field preserve the $C_{3v}$ symmetry, transform as $A_1$ in the Jahn-Teller description, and result in a global shift of the excited-state manifold relative to the ground states. Transverse perturbations ($E_1$ or $E_2$) lower the defect symmetry which results in splitting and mixing the orbital branches. We can express the NV-center interaction with strain and electric field perturbations in the basis {$\ket{E_x}, \ket{E_y}$} as \cite{Maze2011b}:
\begin{equation}
\label{eqn: Eqn 1}
\mathcal{H} = \mathcal{H}_0 + (V_{\textrm{A}_1} - \mathcal{E}_z)I + (V_{\textrm{E}_1} - \mathcal{E}_x)\sigma_{z} + (V_{\textrm{E}_2} - \mathcal{E}_y)\sigma_{x}    
\end{equation}
where $\mathcal{H}_0$ is the unperturbed NV-center ZPL transition energy, $V_{\Gamma}$ are the strain deformation potentials transforming as the irreducible representations of $C_{3v}$, and $\mathcal{E}_i$ are the Stark potentials resulting from electric fields along the $\hat{i}$ direction. Here $\hat{z}$ is along the NV-center symmetry axis and $\hat{x}, \hat{y}$ lie in a plane defined by the three carbon atoms adjacent to the vacancy [Figure \ref{fig:Fig 1}(b)]. By measuring the global shift ($A_1$-type) and splitting ($E$-type) of $\ket{E_x}$ and $\ket{E_y}$ we can decompose the perturbations into different symmetries.

\begin{figure*}
\includegraphics{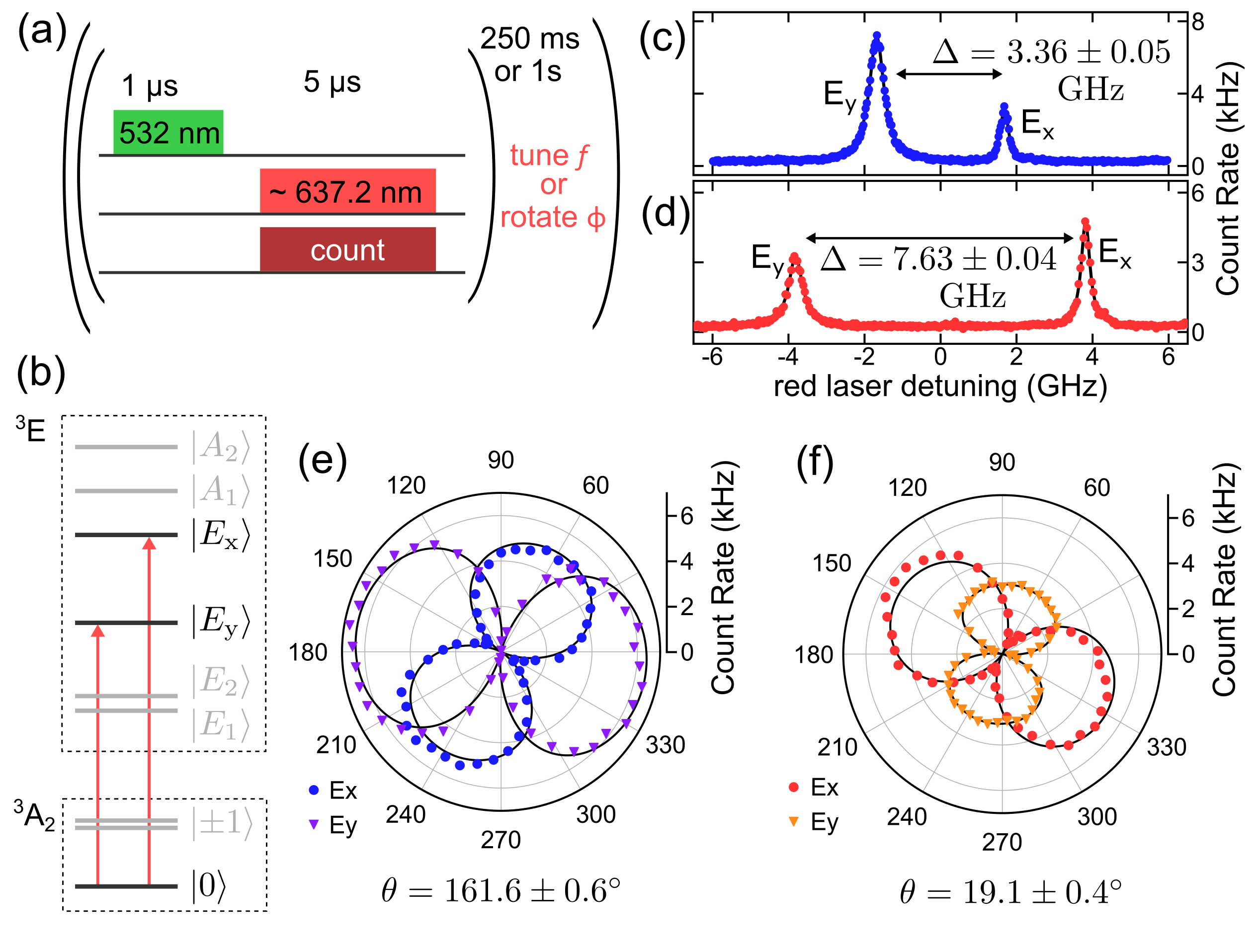}
\caption{\label{fig:Fig 2}(a) Pulse sequence diagram for static strain measurement. Repeated green laser excitation prepares the charge state to $\textrm{NV}^{-}$ and the spin state to $m_s = 0$ while averaging over spectral diffusion. (b) NV center ground and excited-state manifold. $m_s = 0$ states shown in black and $m_s = \pm 1$ shown in grey. (c) PLE frequency sweep of NV-8 and (d) NV-9 and extracted strain splitting $\Delta$ from two Lorentzian fit. Detuning for each defect plotted relative to the absolute center frequency of each. Pulse scheme applied for 250 ms per frequency for PLE frequency sweeps. (e) PLE laser polarization rotation of NV-8 and (f) NV-9. Red laser is tuned onto resonance with either $\ket{E_x}$ or $\ket{E_y}$ for polarization rotation measurements. NV-center dipole rotation $\theta$ extracted from fit given in Appendix \ref{sec:Appendix A}. Pulse scheme applied for 1 s at each angle for PLE polarization rotation.}
\end{figure*}

We begin by characterizing the static strain of each NV center using the pulse scheme shown in Figure \ref{fig:Fig 2}(a). A green laser pulse (100 $\mu$W, 532 nm) initializes the NV-center spin to $\textrm{m}_s = 0$ and charge to $\textrm{NV}^{-}$. A red laser (200 nW, $\sim$637.2 nm) is frequency tuned across the ZPL transitions as we monitor the phonon sideband fluorescence. The frequency-swept PLE response of two example defects in our study, referred to here as NV-8 and NV-9, is shown in Figure \ref{fig:Fig 2}(c,d). We simultaneously monitor the absolute laser frequency using a wavemeter ($\sim$ 1 GHz resolution) and the relative laser detuning with a Fabry-Perot cavity ($\sim$ 10 MHz resolution). Zero detuning for the PLE response of NV-8 and NV-9 is set to the center frequency of $\ket{E_x}$ and $\ket{E_y}$ for each defect. Repeated application of the green laser at each laser frequency causes a sampling over charge trap configurations that results in an inhomogeneously broadened PLE spectrum and peak positions with no net contribution from electric fields. Setting the electric field terms to zero, Equation \ref{eqn: Eqn 1} reduces to:

\begin{equation}
\label{eqn: Eqn 2}
\mathcal{H} = \mathcal{H}_0 + V_{\textrm{A}_1}I + V_{\textrm{E}_1}\sigma_{z} + V_{\textrm{E}_2}\sigma_{x}    
\end{equation}

To find $V_{\textrm{A}_1}$ we extract the absolute center frequency of $\ket{E_x}$ and $\ket{E_y}$ for our 16 defects and calculate the absolute deviation of each defect's center frequency from the mean transition frequency. This procedure can result in a constant offset in the extracted $V_{\textrm{A}_1}$ values if the mean absolute transition frequency of our defects does not correspond exactly to the transition frequency for zero longitudinal strain. We explore the consequences of such an offset in Appendices \ref{sec:Appendix B} and \ref{sec:Appendix D}, finding that such an effect does not influence our conclusions.

The $\ket{E_x} \leftrightarrow \ket{E_y}$ frequency splitting $\Delta$ is determined by the total $E$-type strain, and can be further decomposed into $E_1$ and $E_2$ components by measuring the rotation $\theta$ of the defect's optical dipole away from equilibrium. $\Delta$ and $\theta$ are related to the $E$-type strains by \cite{Lee2016}:

\begin{equation}
\label{eqn: Eqn 3}
\begin{split}
\Delta = 2\sqrt{{V_{\textrm{E}_1}}^2 + {V_{\textrm{E}_2}}^2} \\
\tan(2\theta) = V_{\textrm{E}_2}/V_{\textrm{E}_1}
\end{split}
\end{equation}

Optical dipole rotation measurements for NV-8 and NV-9 are shown in Figure \ref{fig:Fig 2}(e,f). To collect these spectra we first tune the red laser into resonance with $\ket{E_y}$. We apply the repeated green and red pulse scheme [Figure \ref{fig:Fig 2}(a)] and rotate the linear polarization angle $\phi$ of the red laser using a half wave plate. PLE counts are acquired for 1 second at each laser polarization. $\phi = 0$ degrees corresponds to the diamond $\langle 110 \rangle$ axis. We repeat this procedure for the $\ket{E_x}$ peak. We extract $\theta$ by fitting the polarization rotation measurements to the known response \cite{Lee2016} (see Appendix A). Finally, we use $\Delta$ and $\theta$ to solve Equation \ref{eqn: Eqn 2} for $V_{\textrm{E}_1}$ and $V_{\textrm{E}_2}$. The static strain components of all sixteen NV centers, as well as the bond axis groups extracted from the optical dipole rotation fitting (see Appendix \ref{sec:Appendix A}) are recorded in Table \ref{tab:Table 1}. All static strains are reported as absolute values.

\begin{figure*}
\includegraphics[scale=0.95]{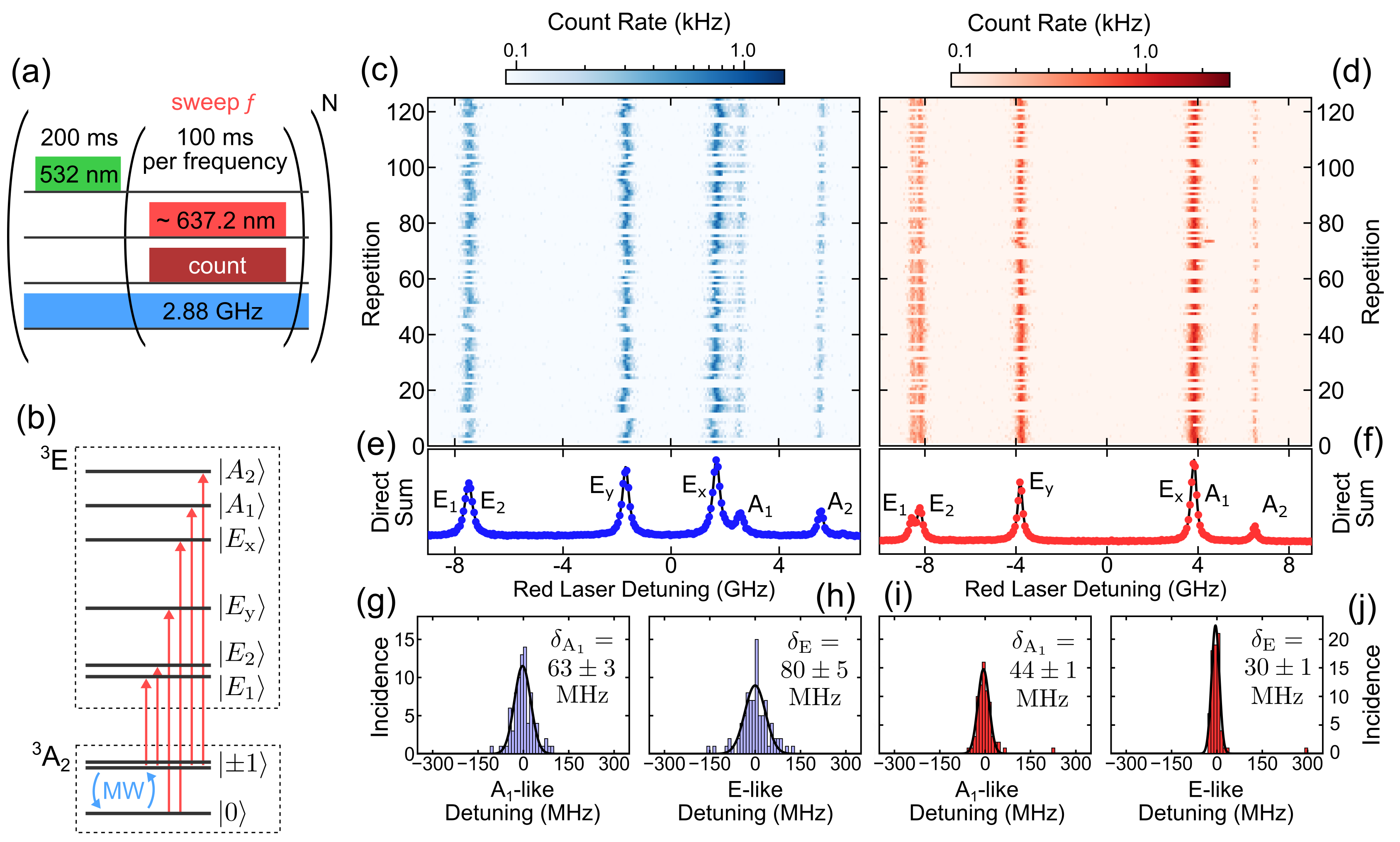}
\caption{\label{fig:Fig 3}(a) Pulse sequence diagram for spectral diffusion measurement. The green pulse before each red laser sweep prepares the charge to $\textrm{NV}^{-}$ and alters local charge trap environment. (b) NV center ground and excited-state manifold. A continuous microwave drive mixes the NV-center spin, allowing PLE detection of all optical transitions. (c) 125 repetitions of frequency-swept PLE on NV-8 and (d) NV-9 showing spectral diffusion on repeated measurement. (e) Direct sum of repeated PLE response for NV-8 and (f) NV-9 with excited state for each transition labelled. (g) Histogram of $A_1$-like and (h) $E$-like spectral diffusion for NV-8 and (i, j) NV-9 extracted from shifting- and splitting-like wandering of $\ket{E_x}$ and $\ket{E_y}$ transitions. Fits are to a Gaussian distribution with the full width at half maximum for each component of spectral diffusion ($\delta_{\textrm{A}_1}$, $\delta_{\textrm{E}}$) reported. Detunings in panels (c-f) are with respect to center frequency of each defect.}
\end{figure*}

Next we quantify the spectral diffusion for each defect. We modify the PLE pulse scheme as shown in Figure \ref{fig:Fig 3}(a). An initial green laser pulse (30 $\mu$W) resets the charge state to $\textrm{NV}^{-}$. The red laser (50 nW) frequency is then swept across the ZPL transitions and PLE counts are recorded. Each measurement of the spectrum follows a single green pulse and therefore corresponds to a single charge trap configuration. Because the $\ket{0} \leftrightarrow \ket{E_{y}}$ transition is known to be non-cycling for some values of transverse strain \cite{Tamarat2008, Batalov2009a}, we apply continuous microwave drive at the ground-state spin transition frequency of 2.88 GHz to sample all NV-center spin configurations [Figure \ref{fig:Fig 3}(b)]. We record the PLE spectrum of each defect 125 times. The spectra for NV-8 and NV-9 are shown in Figure \ref{fig:Fig 3}(c-d) that each show drift of the frequencies on repeated measurement. Photoionization of the NV center during a sweep causes only some of the expected PLE peaks to be detected and unsuccessful preparation of the NV-center charge state to $\textrm{NV}^{-}$ by the green laser results in sweeps with no PLE response. The direct sums of the 125 sweeps for NV-8 and NV-9 are plotted in Figure \ref{fig:Fig 3}(e-f) with the excited state corresponding to each peak labelled.

To quantify the spectral diffusion we first extract the center frequency and splitting of the the averaged $\ket{E_x}$ and $\ket{E_y}$ peaks in Figure \ref{fig:Fig 3}(e,f). Then, for each individual PLE sweep in Figure \ref{fig:Fig 3}(c,d) where both the $\ket{E_x}$ and $\ket{E_y}$ photon counts are above a certain threshold, we extract a set of frequencies for each peak. We remove sweeps with photoionized spectra. We then calculate the $A_1$-like diffusion of each sweep from the shift of the center frequency relative to the averaged center frequency, and the $E$-like diffusion by calculating the change in the $\ket{E_x} \leftrightarrow \ket{E_y}$ splitting relative to the average splitting. The $E$-like splitting results from a combination of $E_1$ and $E_2$-like perturbations, which we are unable to further decompose. We construct histograms of the $A_1$-like and $E$-like diffusion components for each defect and fit with Gaussian distributions. We use the full width at half maximum of the Gaussian fits to quantify the degree of spectral diffusion for each symmetry. The spectral diffusion histograms for NV-8 and NV-9 are shown in Figure \ref{fig:Fig 3}(g-j). The degree of spectral diffusion of each symmetry for each defect is recorded in Table \ref{tab:Table 1}.

\renewcommand{\arraystretch}{0.7}
\begin{table*}
\caption{\label{tab:Table 1}Strain, Spectral Diffusion, Depth from Surface, and Bond Axis Groups}
\begin{ruledtabular}
\begin{tabular}{
    c c c c c c c c c c c
        }
 \multicolumn{1}{c}{defect} & \multicolumn{1}{c}{$V_{\textrm{A}_1}$ (GHz)} & \multicolumn{1}{c}{$V_{\textrm{E}_1}$ (GHz)} & \multicolumn{1}{c}{$V_{\textrm{E}_2}$ (GHz)} & \multicolumn{1}{c}{$\delta_{\textrm{A}_1}$ (MHz)} & \multicolumn{1}{c}{$\delta_{\textrm{E}}$ (MHz)} & \multicolumn{1}{c}{$d$ ($\mu$m)} & \multicolumn{1}{c}{Group} \\ \hline

NV-1	&	11.9	$\pm$	1.0	&	2.28	$\pm$	0.07	&	2.42	$\pm$	0.07	&	207	$\pm$	16	&	92	$\pm$	8	&	33.5	$\pm$	0.3	&	A\\
NV-2	&	4.5	$\pm$	1.3	&	0.37	$\pm$	0.08	&	2.56	$\pm$	0.04	&	83	$\pm$	7	&	168	$\pm$	14	&	22.2	$\pm$	0.3	&	B\\
NV-3	&	3.4	$\pm$	1.3	&	0.30	$\pm$	0.11	&	2.36	$\pm$	0.05	&	91	$\pm$	6	&	68	$\pm$	3	&	35.0	$\pm$	0.4	&	B\\
NV-4	&	23.1	$\pm$	1.0	&	2.44	$\pm$	0.04	&	0.87	$\pm$	0.05	&	120	$\pm$	12	&	147	$\pm$	12	&	33.7	$\pm$	0.3	&	B\\
NV-5	&	2.7	$\pm$	1.6	&	0.62	$\pm$	0.06	&	2.56	$\pm$	0.04	&	90	$\pm$	6	&	79	$\pm$	5	&	26.5	$\pm$	0.3	&	A\\
NV-6	&	3.0	$\pm$	1.5	&	1.63	$\pm$	0.05	&	1.91	$\pm$	0.05	&	132	$\pm$	11	&	56	$\pm$	3	&	20.0	$\pm$	0.3	&	A\\
NV-7	&	4.2	$\pm$	1.5	&	1.53	$\pm$	0.08	&	3.53	$\pm$	0.06	&	69	$\pm$	5	&	50	$\pm$	2	&	15.8	$\pm$	0.3	&	A\\
NV-8	&	7.5	$\pm$	1.6	&	1.34	$\pm$	0.04	&	1.01	$\pm$	0.04	&	63	$\pm$	3	&	80	$\pm$	5	&	13.8	$\pm$	0.3	&	B\\
NV-9	&	8.1	$\pm$	1.2	&	3.00	$\pm$	0.05	&	2.35	$\pm$	0.06	&	44	$\pm$	1	&	30	$\pm$	1	&	9.1	$\pm$	0.3	&	A\\
NV-10	&	10.3	$\pm$	1.3	&	2.31	$\pm$	0.07	&	2.41	$\pm$	0.07	&	66	$\pm$	3	&	22	$\pm$	1	&	7.1	$\pm$	0.3	&	A\\
NV-11	&	6.3	$\pm$	1.2	&	0.64	$\pm$	0.09	&	2.42	$\pm$	0.06	&	81	$\pm$	5	&	46	$\pm$	3	&	12.5	$\pm$	0.3	&	B\\
NV-12	&	4.6	$\pm$	1.2	&	1.22	$\pm$	0.09	&	1.56	$\pm$	0.09	&	223	$\pm$	19	&	89	$\pm$	4	&	30.5	$\pm$	0.3	&	A\\
NV-13	&	0.3	$\pm$	1.1	&	1.17	$\pm$	0.05	&	0.81	$\pm$	0.05	&	260	$\pm$	32	&	48	$\pm$	4	&	22.1	$\pm$	0.3	&	A\\
NV-14	&	6.7	$\pm$	0.9	&	0.16	$\pm$	0.11	&	3.14	$\pm$	0.05	&	87	$\pm$	4	&	112	$\pm$	5	&	26.6	$\pm$	0.3	&	B\\
NV-15	&	1.5	$\pm$	0.8	&	1.94	$\pm$	0.07	&	2.20	$\pm$	0.07	&	172	$\pm$	13	&	54	$\pm$	2	&	10.4	$\pm$	0.3	&	B\\
NV-16	&	8.0	$\pm$	1.2	&	2.97	$\pm$	0.04	&	0.18	$\pm$	0.02	&	132	$\pm$	13	&	23	$\pm$	2	&	10.8	$\pm$	0.3	&	A\\

\end{tabular}
\end{ruledtabular}
\end{table*}

Several noteworthy features stand out when strain and spectral diffusion components are decomposed. For example, from Figure \ref{fig:Fig 3}(g-j) we can see that NV-8 has overall more spectral diffusion than NV-9. Additionally, NV-8 has spectral diffusion dominated by $E$-like perturbations, whereas NV-9 has a diffusion that is more $A_1$-like. These provide a sharp contrast with the static strain components reported for these two defects (see Table \ref{tab:Table 1}). NV-8 has overall lower strain than NV-9, and the strain is primarily more $A_1$-like than $E$-like. Despite NV-9 having larger $E$-like strain, its spectral diffusion is dominated by $A_1$-like contributions.

\section{Correlations}

\begin{figure}[h!]
\includegraphics[scale=0.9]{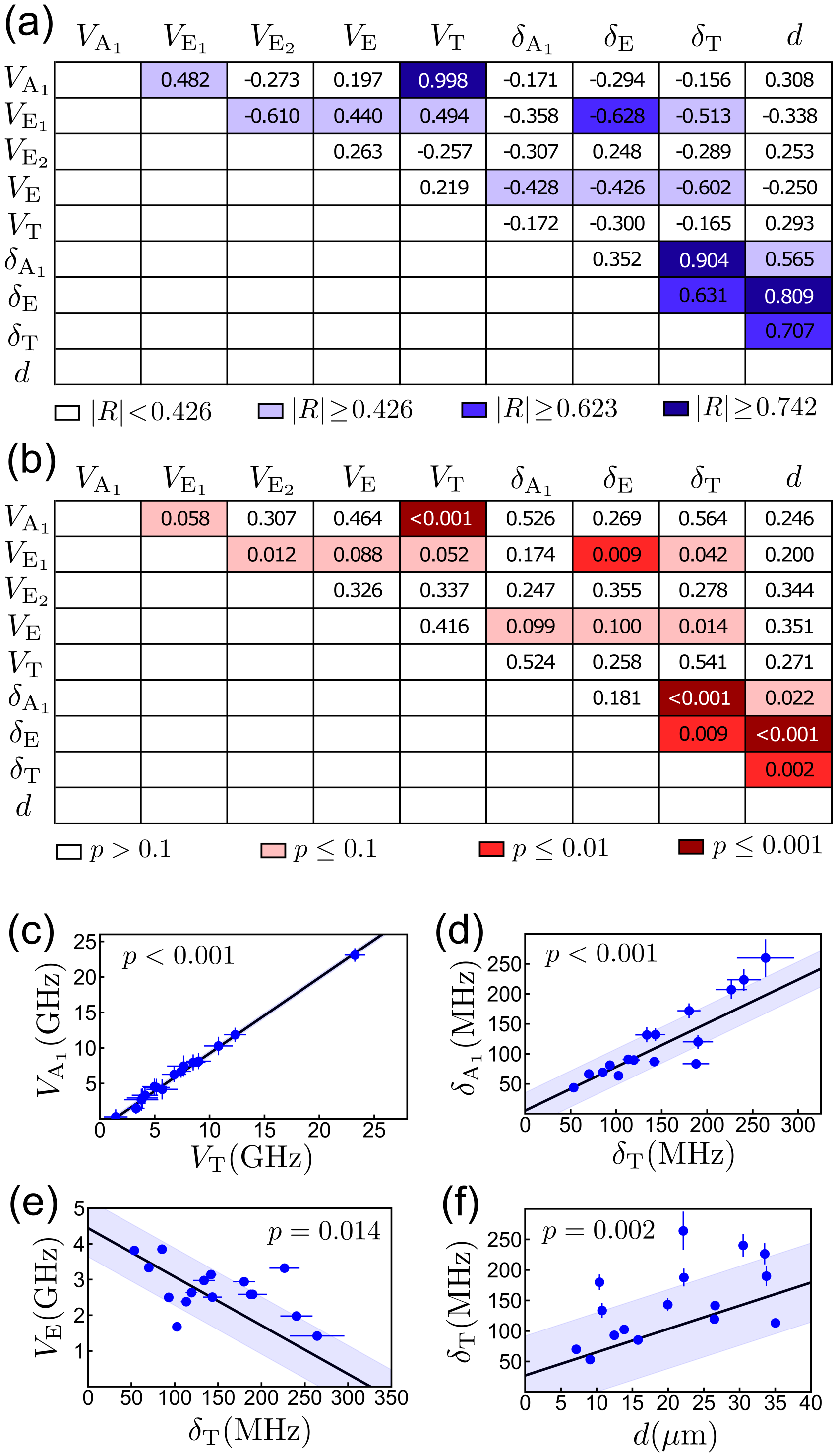}
\caption{\label{fig:Fig 4} (a) Calculated correlation coefficients $R$ and (b) $p$-values for sets of the measured physical parameters of all sixteen NV centers. (c) $V_{\textrm{A}_1}$ versus $V_{\textrm{T}}$ and (d) $\delta_{\textrm{A}_1}$ versus $\delta_{\textrm{T}}$ show strong positive correlation. (e) $V_{\textrm{E}}$ versus $\delta_{\textrm{T}}$ shows negative correlation. (f) $\delta_{\textrm{T}}$ versus $d$ shows positive correlation. Least-squares linear fits and one standard deviation prediction intervals are shown as guides to the eye. Color scale cutoffs for $p$-values are given as orders of magnitude, and cutoffs for $R$-values are the corresponding correlation coefficients for those $p$-values.}
\end{figure}

Using this data set, we now investigate underlying physical relationships between measured parameters by looking at correlations. We consider the components of static strain ($V_{\textrm{A}_1}$, $V_{\textrm{E}_1}$, $V_{\textrm{E}_2}$), components of spectral diffusion ($\delta_{\textrm{A}_1}$, $\delta_{\textrm{E}}$), and measured depth from the surface $d$, as well as the overall $E$-like static strain $V_\textrm{E} = \sqrt{{V_{\textrm{E}_1}}^2 + {V_{\textrm{E}_2}}^2}$, overall strain $V_\textrm{T} = \sqrt{{V_{\textrm{A}_1}}^2 + {V_\textrm{E}}^2}$, and overall spectral diffusion $\delta_\textrm{T} = \sqrt{{\delta_{\textrm{A}_1}}^2 + {\delta_{\textrm{E}}}^2}$. We calculate the weighted Pearson correlation coefficients $R$, which measure the degree of linear correlation between two sample data sets, and the two-tailed $p$-values which give the probability of finding a correlation as extreme as the computed value, assuming that the populations from which the sample data are drawn are uncorrelated. The details of these calculations are given in Appendix \ref{sec:Appendix B}. We report the calculated $R$ and $p$-values in Figure \ref{fig:Fig 4}(a,b), including a color scale indicating the degree of the correlation and significance. Because our sample data set size is only 16, we also consider the influence of outlier data points by performing robustness analysis of the calculated $R$ and $p$-values in Appendix \ref{sec:Appendix C}. We find that small sample size effects do not play a significant role in our analysis.

The most strongly correlated sets of parameters are $V_{\textrm{A}_1}$ with $V_\textrm{T}$ and $\delta_{\textrm{A}_1}$ with $\delta_\textrm{T}$ as shown in Figure \ref{fig:Fig 4}(c,d). Some correlation is expected since the $V_\textrm{T}$ and $\delta_\textrm{T}$ both depend directly on their own $A_1$-like components, but the higher degree of correlation with $A_1$ components rather than $E$-like components reveals that both static strain and spectral diffusion are dominated by $A_1$-type perturbations. This is in agreement with a previous study of applied strain-orbital coupling in NV centers showing similar $A_1$-dominant behavior \cite{Lee2016}. Based on first principles calculations, the spectral shifts induced by electric fields are also expected to be larger for $A_1$-like electric fields \cite{Maze2011b}. 

Notable also is the correlation between $V_\textrm{E}$ and $\delta_\textrm{T}$ shown in Figure \ref{fig:Fig 4}(e). The negative slope of the linear dependence and the degree of correlation suggests that for NV centers with larger off-axis strain, the effects of spectral diffusion are diminished. This agrees well with previous first-principles calculations of NV-center response to both transverse strain and transverse electric fields, where it was found that transverse electric field contributions that are orthogonal to the transverse strain on the NV-center contribute quadratically rather than linearly to the NV center energy \cite{Maze2011b}. 

One surprising result is found in Figure \ref{fig:Fig 4}(f) when total spectral diffusion $\delta_\textrm{T}$ is plotted against depth from surface $d$. The correlation of these parameters and the positive linear trend suggests that the spectral diffusion increases with NV-center depth. We note that all NV centers in our study are deep, with the shallowest being at $7.1 \pm 0.3$ $\mu$m from the surface. Therefore, we do not expect charge traps at the diamond surface to affect the observed spectral diffusion. One potential explanation is that the green laser excites a greater number of charge trap sites with increasing depth because of optical aberration to the confocal laser excitation spot. To put this in perspective, a single electron charge in the diamond bulk produces an electric field of roughly 1 kV/m at a distance of 500 nm, which corresponds to a spectral shift of order 10 MHz \cite{Tamarat2006, Maze2011b}. A second potential explanation for observing increased spectral diffusion at increased depth is if a depth-related strain profile leads to protection from spectral diffusion nearer to the surface where strain may be larger. However, we find significantly stronger correlation of spectral diffusion with depth than we do of strain with depth. Additionally, $\delta_{\textrm{A}_1}$-like diffusion should not be affected by strain. The significant correlation of $\delta_{\textrm{A}_1}$ with depth suggests that depth-related aberrations are the dominant mechanism.

\section{Discussion}

We have performed symmetry decomposed measurements of spectral diffusion in NV centers using a group theory informed approach to PLE spectroscopy. We find three interesting sets of correlated parameters. First, we find that the NV-center strain and spectral diffusion are each dominated by perturbations along the NV-center symmetry axis. Second, we find that off-axis strain on the NV center provides some degree of protection from spectral diffusion. This result contrasts with the result of a previous study of this relationship in thin diamond membranes \cite{Ruf2019}, though we note that for the subset of NV centers in that study with comparable distance to the diamond surface (where surface charge effects can be ignored) the general pattern of less spectral diffusion for increasing transverse strain appears to hold. Third, we find a surprising increase of spectral diffusion with increasing depth from the surface. This result suggests a sweet spot in depth for ideal NV-center optical linewidth: far enough from the surface to eliminate sensitivity to surface charge fluctuations but shallow enough to limit focusing aberrations.

Our symmetry decomposed technique for studying strain and spectral diffusion can aid in mitigating the defect-to-defect variations in optical frequencies and the wandering of the optical frequencies which arise from different physical sources. Overcoming both requires separate, but potentially complementary strategies. Depending on the application, our technique suggests a preference for NV centers with large transverse strain that are more likely to have reduced spectral diffusion. This can be helpful for quickly identifying high quality NV centers. Decomposition of spectral diffusion can be used to understand the origin and spatial distribution of charge trap sites, particularly those responsible for the significant spectral diffusion seen in patterned structures \cite{Faraon2012, Riedel2017, Ruf2019}, as well as for engineering better frequency stability of implanted NV centers \cite{vanDam2019, Kasperczyk2020, Yurgens2021, Chakravarthi2021}. The combination of symmetry decomposed spectral diffusion measurements alongside charge state stability measurements of near-surface NV centers \cite{Bluvstein2019} can provide additional information about the microscopic physics of surface charge traps. Our technique will aid in investigating mechanically driven continuous orbital dynamical decoupling schemes \cite{Chen2018} to actively decouple a defect's transition energy from spectral diffusion, since defects dominated by $E$-type diffusion can in principle be stabilized against such perturbations. 

\section{Acknowledgements}

We thank Kenneth Lee for insightful and thorough discussions regarding NV-center polarization dependent optical transitions. We acknowledge support from the Office of Naval Research under Grant No. N00014-21-1-2614. H.F.H. Cheung acknowledges funding from the Department of Energy Office of Science, Basic Energy Sciences grant DE-SC0019250. Device fabrication was performed in part at the Cornell Nanoscale Facility, a member of the National Nanotechnology Coordinated Infrastructure (NNCI), which is supported by the NSF (NNCI-2025233), and at the Cornell Center for Materials Research Shared Facilities which are supported through the NSF MRSEC program (Grant No. DMR-1719875).

\clearpage

\begin{appendix}

\counterwithin{figure}{section}
\setcounter{figure}{0}

\section{Polarization dependence of NV-center optical transitions}\label{sec:Appendix A}


\begin{figure}
\includegraphics{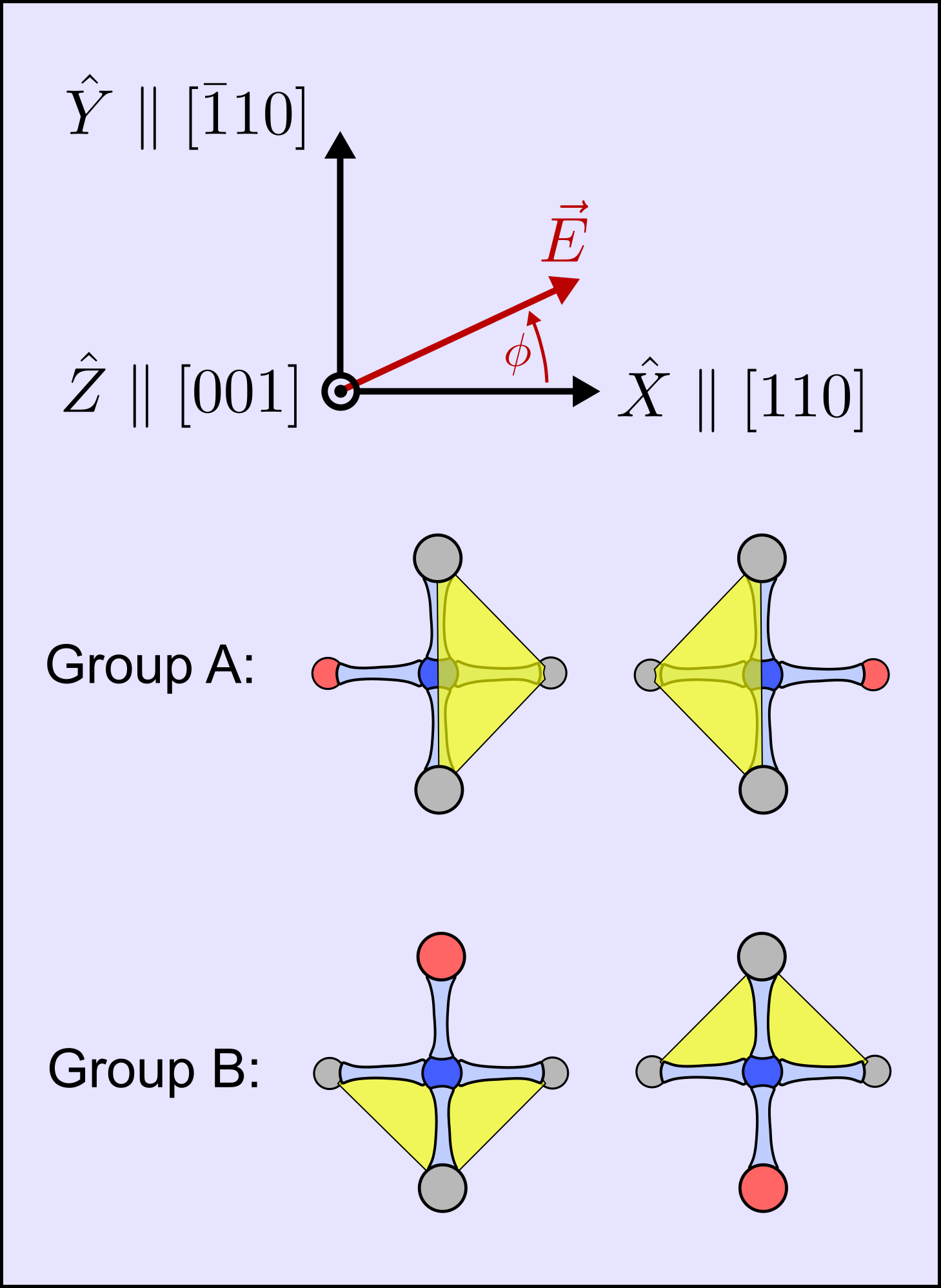}
\caption{\label{fig:Fig S1}Sample coordinate system used for polarization dependence of optical transitions. Diamond substrate (light blue) has $\langle 100 \rangle$ oriented faces and $\langle 110 \rangle$ oriented edges. Laser polarization direction indicated by $\phi$. Two groups of optical responses are observed depending on the orientation of the NV-center bond axis direction. Group A is given by $\hat{z} = \left[ \bar{1} \bar{1} \bar{1} \right]$ and $\hat{z} = \left[ 1 1 \bar{1} \right]$, and Group B is given by $\hat{z} = \left[ \bar{1} 1 1 \right]$ and $\hat{z} = \left[ 1 \bar{1} 1 \right]$. Their projections as viewed along the diamond $\left[ 0 0 \bar{1} \right]$ (the plane of the laser polarization) are shown. Dipole moments responsible for optical transitions lie in the plane (yellow) perpendicular to the bond between the vacancy (blue circles) and nitrogen atom (red circles), with $\hat{x}$ chosen to lie in a $C_{3v}$ reflection plane (a plane containing the nitrogen, vacancy, and one carbon atom).}
\end{figure}

Quantifying the two transverse strain potentials $V_{\textrm{E}_1}$ and $V_{\textrm{E}_2}$ requires determination of the $\ket{E_x} \leftrightarrow \ket{E_y}$ splitting $\Delta$ and the rotation $\theta$ of the optical dipole away from equilibrium. Extraction of the splitting is straightforward, as is shown in Figure \ref{fig:Fig 2}. To determine the dipole rotation we follow the method of Lee et al. \cite{Lee2016} with a few noteworthy modifications.

We define the coordinate system of our diamond sample as shown in Figure \ref{fig:Fig S1}: $\hat{X} = \left[ 1 1 0 \right]$, $\hat{Y} = \left[ \bar{1} 1 0 \right]$, and $\hat{Z} = \left[ 0 0 1 \right]$. The electric field of the resonant laser responsible for driving optical transitions can be expressed as $\vec{E} = E_0 \hat{r}$ where $\hat{r} = \cos (\phi) \hat{X} + \sin (\phi) \hat{Y} $. Rewriting this in the coordinate system of our diamond gives:

\begin{equation}
\label{eqn: Eqn S1}
\hat{r} = \frac{\cos (\phi)}{\sqrt{2}} \left[ 1 1 0 \right] + \frac{\sin (\phi)}{\sqrt{2}} \left[ \bar{1} 1 0 \right]
\end{equation}

Optical transitions are driven by the interaction of the NV-center dipole moment $\vec{d}$ with the electric field of the resonant laser $\vec{E}$ via the electric dipole interaction $H_E = - \vec{d} \cdot \vec{E}$. The transverse components of the electric dipole moment $d_x$ and $d_y$ couple the ground and excited-state manifolds. In the basis $\{ \ket{0},\ket{E_x},\ket{E_y} \}$ composed of $m_s=0$ states and in the absence of strain, the transverse electric dipole moment components are given by:

\begin{equation}
\label{eqn: Eqn S2}
\begin{split}
d_x = 
\begin{pmatrix}
0 & 0 & d_{\perp}\\
0 & \widetilde{d_{\perp}} & 0\\
d_{\perp} & 0 & \widetilde{d_{\perp}}
\end{pmatrix} \\
d_y = 
\begin{pmatrix}
0 & d_{\perp} & 0\\
d_{\perp} & 0 & \widetilde{d_{\perp}}\\
0 & \widetilde{d_{\perp}} & 0
\end{pmatrix}
\end{split}
\end{equation}

$V_{\textrm{E}_2}$ introduces cross terms into Equation \ref{eqn: Eqn 2} of the main text, and the eigenstates are changed. The result of this interaction is that the transverse dipole moments are rotated in the strain field as follows:

\begin{equation}
\label{eqn: Eqn S3}
\begin{split}
\widetilde{d_x} & = \cos(\theta) d_x + \sin(\theta) d_y \\
\widetilde{d_y} & = -\sin(\theta) d_x + \cos(\theta) d_y
\end{split}
\end{equation}
where $2\theta$ is the St{\"u}ckelberg angle given by $\tan(2\theta) = V_{\textrm{E}_2}/V_{\textrm{E}_1}$

Before calculating the NV-center optical response to laser polarization it is useful to remark on the orientation of NV centers in diamond. NV centers can form along any of the $\langle 111 \rangle$ bond axis directions. For our analysis we choose the NV-center $\hat{z}$-axis to point along the direction of the nitrogen-vacancy bond, and $\hat{x}$ to lie in a $C_{3v}$ reflection plane which is a plane containing the nitrogen, vacancy, and one carbon atom. When excited by a laser propagating along the $\left( 100 \right)$ direction, the optical response of individual NV centers in $\left( 100 \right)$ oriented diamond can be classified into two families of responses. NV centers in Group A have $\hat{z} = \left[ \bar{1} \bar{1} \bar{1} \right]$ ($\hat{x} = \left[ 1 1 \bar{2} \right]$) and $\hat{z} = \left[ 1 1 \bar{1} \right]$ ($\hat{x} = \left[ \bar{1} \bar{1} \bar{2} \right]$). NV centers in Group B have $\hat{z} = \left[ \bar{1} 1 1 \right]$ ($\hat{x} = \left[ 1 \bar{1} 2 \right]$) and $\hat{z} = \left[ 1 \bar{1} 1 \right]$ ($\hat{x} = \left[ \bar{1} 1 2 \right]$). The projections of Group A and Group B NV centers into the $\left( 100 \right)$ plane of the laser electric field polarization is shown in Figure \ref{fig:Fig S1}. The choice of $\hat{z}$ to point from the vacancy to the nitrogen is arbitrary; making the opposite choice results in modified fitting equations that correspond to values of $\theta$ which produce an overall minus sign when $\tan{\theta}$ is evaluated used in Equation \ref{eqn: Eqn 3} to determine $V_{\textrm{E}_1}$ and $V_{\textrm{E}_2}$. Thus, we are only able to reliably determine the magnitude of these strain potentials.

The absorption intensity of the optical transitions is proportional to ${| \vec{d} \cdot \vec{E} |}^2$. For Group A the normalized absorption intensities $N_{E_x}\left( \phi \right)$ and $N_{E_y}\left( \phi \right)$ are given by: 

\begin{equation}
\label{eqn: Eqn S4}
\begin{split}
N_{E_x}\left( \phi \right) & = {\left( \frac{\cos (\theta) \cos(\phi)}{\sqrt{3}} - \sin (\theta) \sin (\phi) \right)}^2 \\
N_{E_y}\left( \phi \right) & = {\left( \frac{\sin (\theta) \cos(\phi)}{\sqrt{3}} - \cos (\theta) \sin (\phi)  \right)}^2
\end{split}
\end{equation}

and for Group B:

\begin{equation}
\label{eqn: Eqn S5}
\begin{split}
N_{E_x}\left( \phi \right) & = {\left( \frac{\cos (\theta) \sin(\phi)}{\sqrt{3}} - \sin (\theta) \cos (\phi) \right)}^2 \\
N_{E_y}\left( \phi \right) & = {\left( \frac{\sin (\theta) \sin(\phi)}{\sqrt{3}} - \cos (\theta) \cos (\phi)  \right)}^2
\end{split}
\end{equation}

The resonant laser power used for our polarization rotation measurements ($\sim 200$ nW) is sufficient to saturate the optical transitions. We model the optical saturation as:

\begin{equation}
\label{eqn: Eqn S6}
I = \left( \frac{1}{1 + \frac{P_{\textrm{sat}}}{P_{\textrm{eff}}}} \right)
\end{equation}
where $P_{\textrm{sat}}$ is the saturation power and $P_{\textrm{eff}}$ is the effective laser power at the NV center. Combining our saturation model with the normalized intensities yields the full optical response functions:

\begin{equation}
\label{eqn: Eqn S7}
\begin{split}
I_{E_x}\left( \phi \right) & = A_{E_x} \left( \frac{1}{1 + \frac{P_{\textrm{sat},x}}{P_{\textrm{in}} N_{E_x}\left( \phi \right)}} \right) \\
I_{E_y}\left( \phi \right) & = A_{E_y} \left( \frac{1}{1 + \frac{P_{\textrm{sat},y}}{P_{\textrm{in}} N_{E_y}\left( \phi \right)}} \right) 
\end{split}
\end{equation}
where $A_{E_x}$ and $A_{E_y}$ are the amplitudes of the optical responses, $P_{\textrm{sat},x}$ and $P_{\textrm{sat},y}$ are the saturation powers, and $P_{\textrm{in}}$ is the input laser power to the objective.

\begin{figure*}
\includegraphics[scale=0.9]{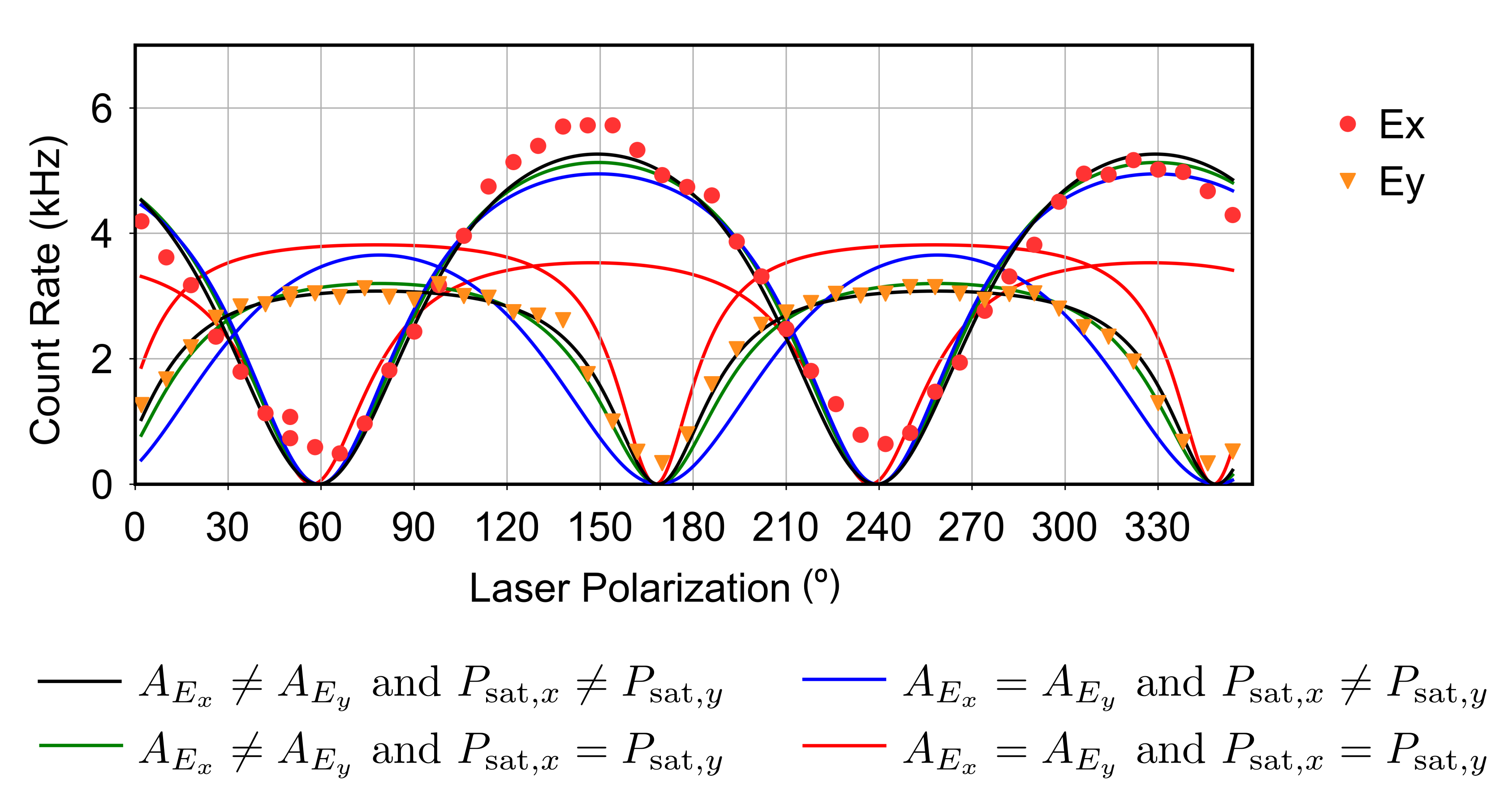}
\caption{\label{fig:Fig S2}Polarization rotation dependent PLE of NV-9. Data is the same as shown in Figure \ref{fig:Fig S2}(f). Four sets of fits are shown. Fit colors indicate the choice of free parameter constraints used.}
\end{figure*}

We do not know a priori which bond axis group each defect in our study occupies. Instead, we fit the polarization rotation dependence of each defect to both the Group A and Group B response and select the set of equations that produces a more reliable fit.

Next, we investigate the justification of free parameters in our model. As an example we consider the polarization response of NV-9, as shown  in Figure \ref{fig:Fig S2}. We simultaneously fit the $I_{E_x}$ and $I_{E_y}$ responses to Equation \ref{eqn: Eqn S7} to extract $\theta$. Exploring the free parameters of our model $A_{E_x}$, $A_{E_y}$, $P_{\textrm{sat},x}$, and $P_{\textrm{sat},y}$, we find the best empirical fit when allowing $\ket{E_x}$ and $\ket{E_y}$ to have independent response amplitudes and saturation powers. Previous studies have suggested different spin flip rates of the $\ket{E_x}$ and $\ket{E_y}$ excited states \cite{Tamarat2008,Batalov2009a}. These differing rates justify the use of independent amplitudes and saturation rates for each optical transition. Regardless, the extracted $\theta$ from fitting does not depend significantly (less than $1 ^{\circ}$) on the choice of free parameter constraints, and error in $\theta$ is predominatnly determined by the $\pm 3 ^{\circ}$ miscut of the diamond substrate.

\begin{figure}
\includegraphics[scale=0.4]{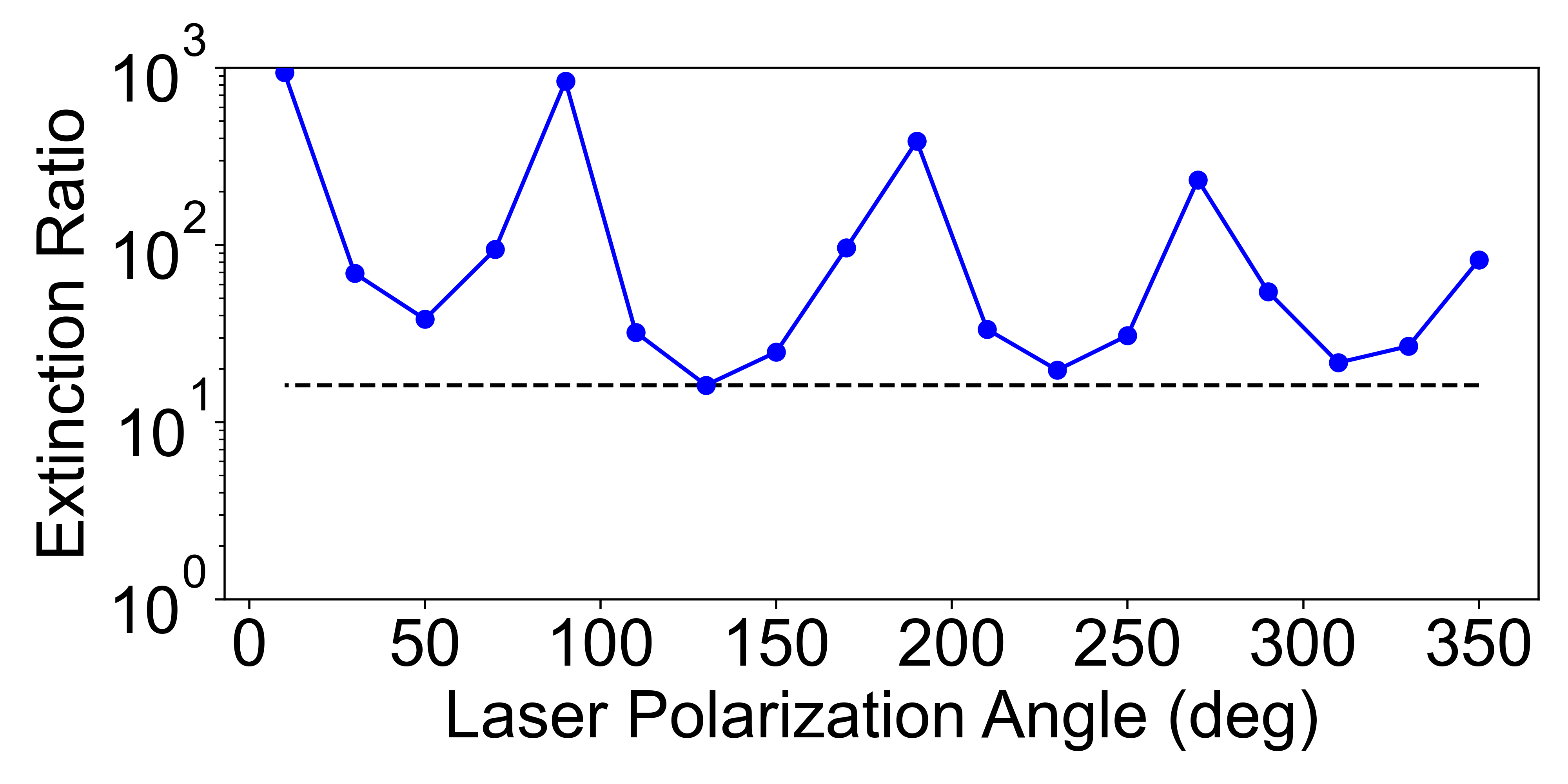}
\caption{\label{fig:Fig S3}Measured laser extinction ratio versus polarization angle at sample. Dotted black line corresponds to the minimum measured extinction ratio of 16:1.}
\end{figure}

Last, we note that Lee et al. included a term in their polarization dependence resulting from laser ellipticity. We record the polarization extinction ratio of our laser as a function of the polarization direction for our setup, reported in Figure \ref{fig:Fig S3}, finding that at worst our extinction is about 16:1. We find that inclusion of our worst-case laser ellipticity to Equation \ref{eqn: Eqn S7} contributes $\ll 1 ^{\circ}$ change to the extracted $\theta$ and so we neglect such considerations for simplicity.

\section{Weighted correlation and $p$-value calculations}\label{sec:Appendix B}


To investigate underlying physical relationships between the collected data sets we calculate the weighted Pearson correlation ($R$) \cite{Pinto_da_Costa_2015} and the corresponding $p$-values for the data sets shown in Table \ref{tab:Table 1} of the main text. Additionally, we consider the calculated $E$-like static strain $V_{\textrm{E}} = \sqrt{{V_{\textrm{E}_1}}^2 + {V_{\textrm{E}_2}}^2}$, overall strain $V_T = \sqrt{{V_{\textrm{A}_1}}^2 + {V_{\textrm{E}}}^2}$, and overall spectral diffusion $\delta_T = \sqrt{{\delta_{\textrm{A}_1}}^2 + {\delta_{\textrm{E}}}^2}$. Our measurement of $V_{\textrm{A}_1}$ assumes that the average absolute center frequency of the $\ket{E_x}$ and $\ket{E_y}$ transitions for our measured set of NV centers corresponds to zero longitudinal strain. Such an assumption leaves the possibility for a constant offset in the reported values of $V_{\textrm{A}_1}$ relative to the values reported in Table \ref{tab:Table 1} of the main text. A constant offset does not influence the calculated correlation or $p$-value for any calculation involving $V_{\textrm{A}_1}$ directly. However, the presence of a constant offset in $V_{\textrm{A}_1}$ could skew the calculation of $V_T$ which could skew the correlations and $p$-values computed using $V_{T}$. The only conclusion in the main text involving $V_T$ is the claim that $A_1$-type strain dominates $E$-type strain in determining $V_T$. We explore this conclusion further in Appendix \ref{sec:Appendix D} and arrive at the same conclusion irrespective of any potential skew from a constant offset in $V_{\textrm{A}_1}$. 

The Pearson correlation coefficient $R_{xy}$ of two sample data sets ($\{ x \}$ and $\{ y \}$) is the ratio of the sample covariance $s_{xy}$ to the square root of the product of the sample variances $s_x$ and $s_y$:

\begin{equation}
\label{eqn: Eqn S8}
R_{xy} = \frac{s_{xy}}{\sqrt{s_x s_y}}
\end{equation}

Each measured data point in our experiment has a corresponding error. We include these measured errors in the correlation analysis by assigning a weight to each $\left( x_i,y_i \right)$ pair. Since in general $\{ x \}$ and $\{ y \}$ do not share a common unit, we use the product of the error bars to assign a single weight to each $\left( x_i,y_i \right)$ pair. The weights are then given by:

\begin{equation}
\label{eqn: Eqn S9}
w_i = \frac{1}{\sigma_{x_i} \sigma_{y_i}}
\end{equation}
where $\sigma_{x_i}$ and $\sigma_{y_i}$ are the error bars of the point $\left( x_i,y_i \right)$. Using these weights we calculate the weighted means $m_x$ and $m_y$ of $\{ x \}$ and $\{ y \}$:

\begin{equation}
\label{eqn: Eqn S10}
m_x = \frac{\sum_{i} w_i x_i}{\sum_{i} w_i} \quad \textrm{and} \quad m_y = \frac{\sum_{i} w_i y_i}{\sum_{i} w_i}
\end{equation}
and using the weighted means we calculate the weighted variances $s_x$ and $s_y$ as:

\begin{equation}
\label{eqn: Eqn S11}
s_x = \frac{\sum_{i} w_i \left( x_i - m_x \right)^2}{\sum_{i} w_i} \quad \textrm{and} \quad s_y = \frac{\sum_{i} w_i \left( y_i - m_y \right)^2}{\sum_{i} w_i}
\end{equation}
and the weighted covariance $s_{xy}$ as:

\begin{equation}
\label{eqn: Eqn S12}
s_{xy} = \frac{\sum_{i} w_i \left( x_i - m_x \right)  \left( y_i - m_y \right)}{\sum_{i} w_i}
\end{equation}
Finally, we use the weighted variances and covariance to calculate the weighted Pearson correlation coefficient $R_{xy}$. The weighted correlations are given in Figure \ref{fig:Fig 4}(a) of the main text.

We also consider the statistical significance of finding such a correlation for each pair of data sets. The statistical significance is given by the $p$-value, valued between 0 and 1. The $p$-value gives the likelihood of drawing a sample data set from uncorrelated variables and finding a correlation at least as strong as the one that was measured. To calculate $p$, we begin by expressing the probability density function of the sample correlation coefficient $f \left( R \right)$ as:

\begin{equation}
\label{eqn: Eqn S13}
f \left( R \right) = \frac{\left( 1 - R^2 \right)^{n/2 - 2}}{B \left( \frac{1}{2} , \frac{n}{2} - 1 \right)}
\end{equation}
where $n$ is the number of points in each sample set (in our experiment $n = 16$) and $B \left( j,k \right)$ is the Beta function given by:

\begin{equation}
\label{eqn: Eqn S14}
B \left( j,k \right) = \int_{0}^{1} u^{j-1} \left( 1-u \right)^{k-1} du
\end{equation}
This probability density function corresponds to the case where the variables are uncorrelated. Calculating the $p$-value using this distribution tests the likelihood of drawing a sample from uncorrelated variables with a correlation as significant as the one measured, as stated above. Since the correlation coefficient $R$ can range from -1 to 1, we consider a two-tailed significance test which tests the likelihood of finding $\lvert R \rvert$ at least as extreme as the one measured. The calculated $p$-value is given by the fraction of the area under the probability density function from $\lvert R \rvert$ to 1:

\begin{equation}
\label{eqn: Eqn S15}
p \left( R \right) = \frac{\int_{ \lvert R \rvert }^{1} f \left( R \right)}{\int_{0}^{1} f \left( R \right)}
\end{equation}
The calculated $p$-values are reported in Table \ref{tab:Table 1} of the main text.


\section{Robustness of statistical results}\label{sec:Appendix C}

\begin{figure*}
\includegraphics[scale=0.8]{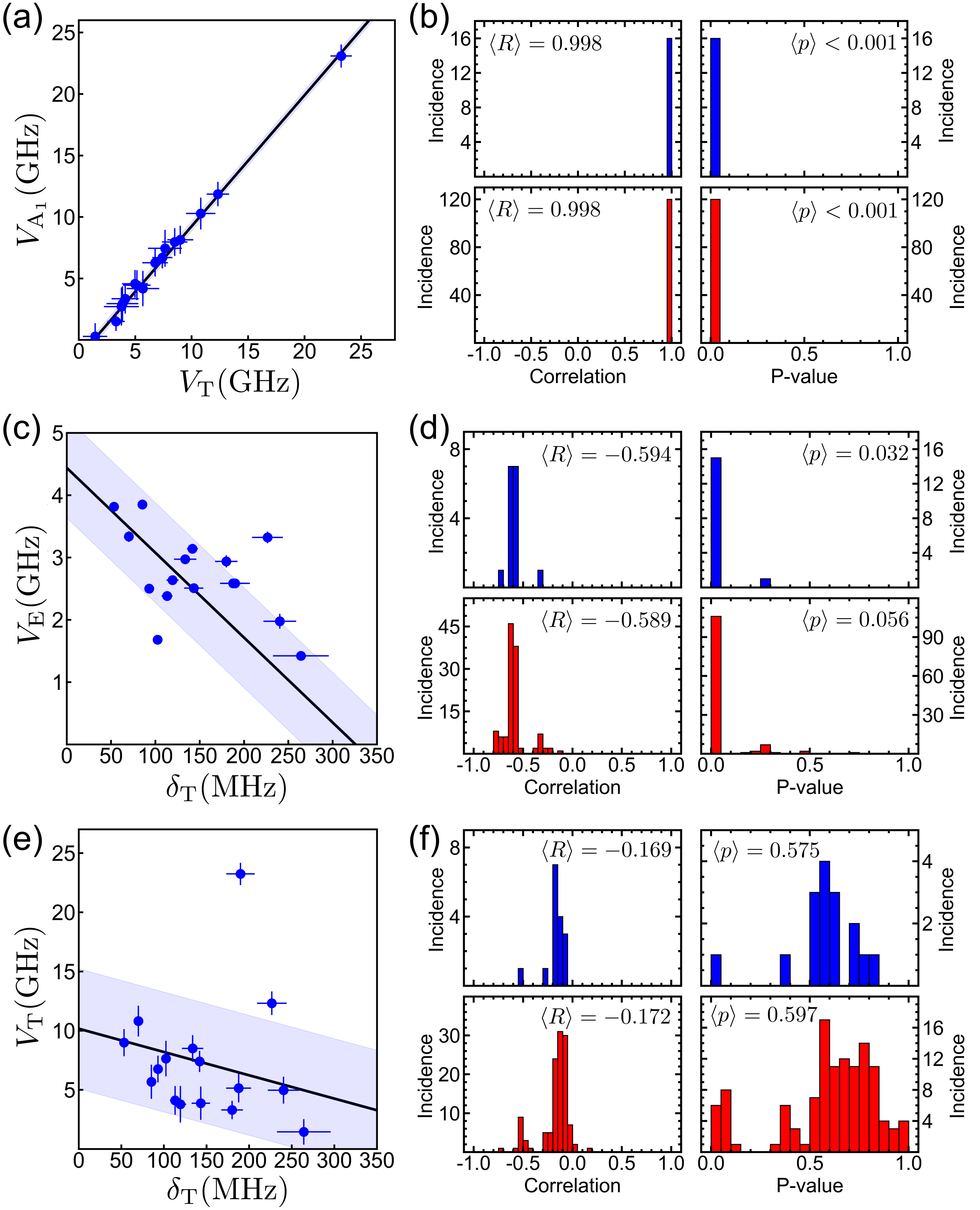}
\caption{\label{fig:Fig S4}Scatter plots of (a) $V_{\textrm{A}_1}$ versus $V_{\textrm{T}}$, (c) $V_{\textrm{E}}$ versus $\delta_{\textrm{T}}$, (e) $V_{\textrm{T}}$ versus $\delta_{\textrm{T}}$. (b,d,f) Calculated correlation and $p$-value for these data sets when removing one data point (blue) or two data points (red) from the analysis, and average value of the calculated $R$ and $p$ for each. Least-squares linear fits and one standard deviation prediction intervals in panels (a,c,e) are provided as guides to the eye.}
\end{figure*}

Since the number of measured defects in our study is relatively small, we perform robustness testing to ensure that the correlations we find are not dominated by outliers. We consider three example cases: the strongly correlated set $\{ V_{\textrm{A}_1} , V_{\textrm{T}} \}$, the weakly correlated set $\{ V_{\textrm{E}} , \delta_{\textrm{T}} \}$, and the uncorrelated set $\{ V_{\textrm{T}} , \delta_{\textrm{T}} \}$. We calculate the $R$ and $p$-values for each pair when either one or two $\{ x_i,y_i \}$ pairs are removed from the data sets. The results of this analysis are shown in Figure \ref{fig:Fig S4}.

\begin{figure*}
\includegraphics[scale = 0.95]{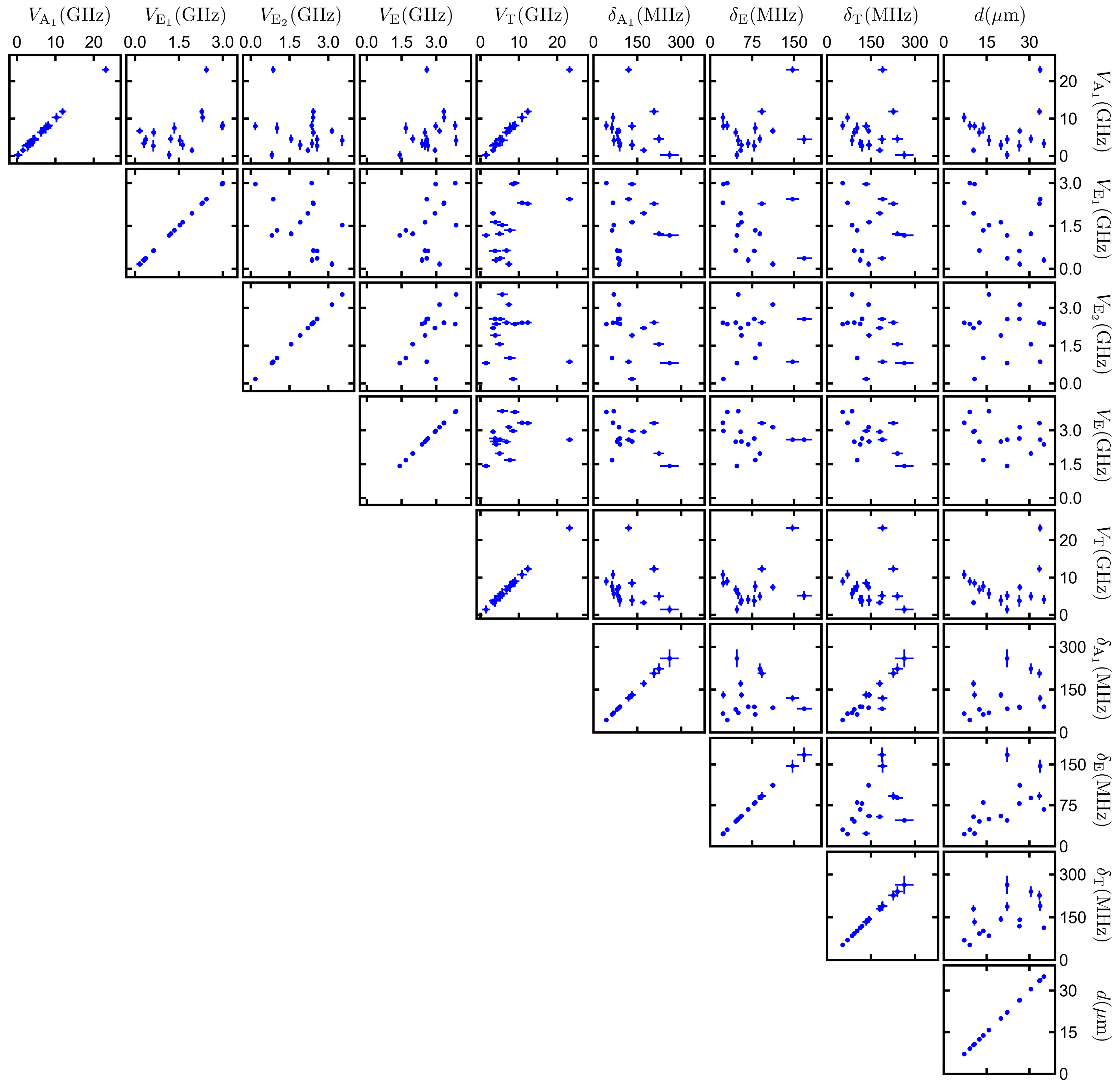}
\caption{\label{fig:Fig S5}Scatter plot matrix of all measured data sets in the experiment. Diagonal terms are included for ease of visualizing the error bars on each data point. Significantly correlated data sets like $\{ V_{\textrm{A}_1} , V_{\textrm{T}} \}$ and $\{ \delta_{\textrm{A}_1} , \delta_{\textrm{T}} \}$ are clearly visible.}
\end{figure*}

For the strongly correlated case of $V_{\textrm{A}_1}$ versus $V_{\textrm{T}}$ there is a clear positive linear relationship. When either one or two data points are removed from the analysis the resulting correlation and $p$-value always results in strong correlation. For the weakly correlated case of $V_{\textrm{E}}$ versus $\delta_{\textrm{T}}$ the scatter plot shows a negative linear relationship. The calculated correlation and $p$-value with one or two data points removed generally show correlation, with average values for $R$ and $p$ comparable to those calculated for the full data sets. The uncorrelated case of $V_{\textrm{T}}$ versus $\delta_{\textrm{T}}$ shows a weak negative linear relationship in the scatter plot. The calculated correlation and $p$-value for these data sets when removing one or two points typically results in no meaningful correlation. The exclusion of the data point which is the largest outlier results in a fraction ($\sim 11 \% $) of cases where the calculated $p$-value is less than 0.10. However, there is no justification to ignore this data point from the analysis rather than any other. On average the $R$ and $p$ values are comparable to those calculated for the full set. For completeness, we record the full scatter plot matrix of all measured data sets in Figure \ref{fig:Fig S5}. Diagonal terms are included in order to demonstrate the relative degree of error bars in each data set.

\section{Dominance of $A_1$-like strain and spectral diffusion}\label{sec:Appendix D}

Both total strain and total spectral diffusion are more correlated with $A_1$-like effects rather than $E$-like effects as shown in the main text. Here, we present additional evidence of this conclusion. Our method for recording $V_{\textrm{A}_1}$ is to first record the absolute center frequency of the $\ket{E_x}$ and $\ket{E_y}$ transitions for each of our sixteen defects and use this information to calculate the absolute deviation of each defect from the mean. The mean of the absolute center frequencies of our sample could differ from the frequency that corresponds to zero longitudinal strain by a constant offset. Such an constant does not affect correlations or $p$-values calculated for quantities involving $V_{\textrm{A}_1}$ directly, but could provide a skew to the calculated $V_{\textrm{T}}$ and subsequent skew to correlations calculated using $V_{\textrm{T}}$, as outlined in Appendix \ref{sec:Appendix B}.

\begin{figure}
\includegraphics[scale=0.9]{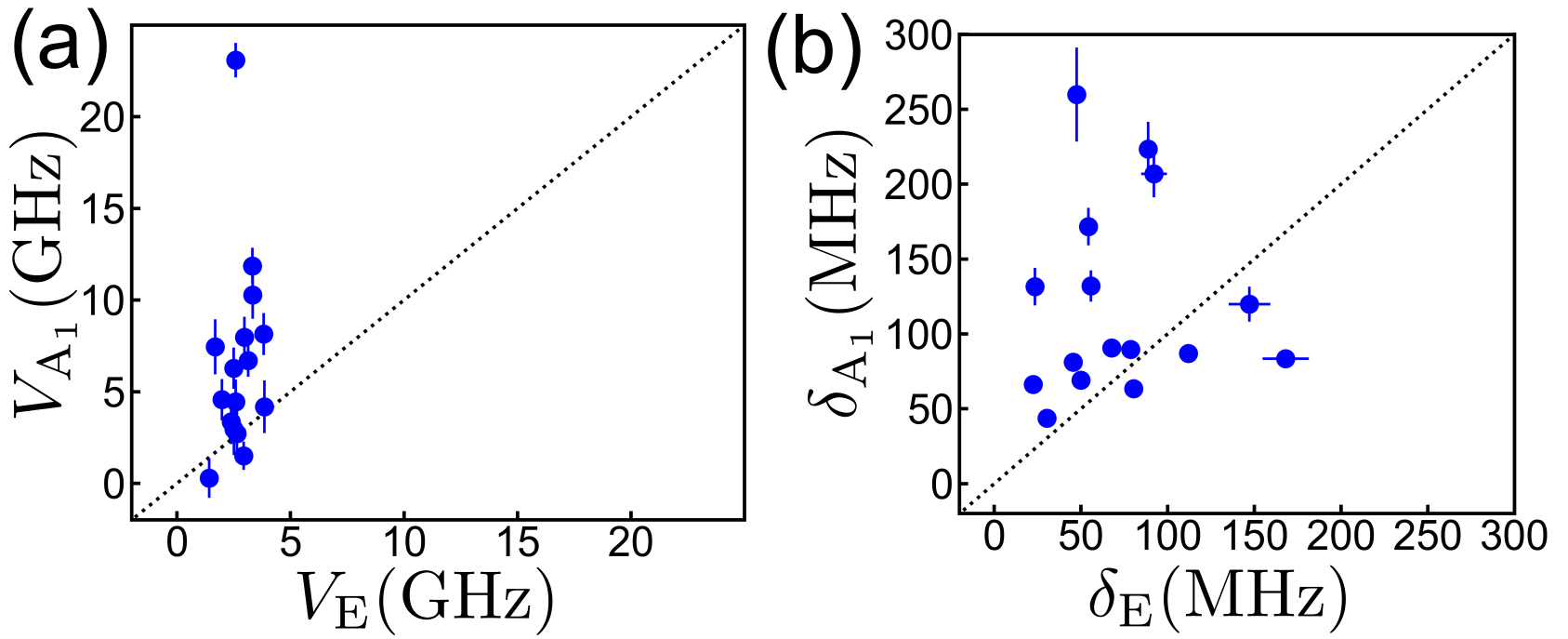}
\caption{\label{fig:Fig S6}(a) $V_{\textrm{A}_1}$ versus $V_{\textrm{E}}$ and (b) $\delta_{\textrm{A}_1}$ versus $\delta_{\textrm{E}}$ each show more defects affected primarily by $A_1$-like perturbation than defects affected primarily by $E$-like. Dotted lines in each panel have a slope of 1.}
\end{figure}

Rather than rely solely on the correlations calculated using $V_{\textrm{T}}$, we can draw the same conclusion by looking at only the components. In Figure \ref{fig:Fig S6} we plot $V_{\textrm{A}_1}$ versus $V_{\textrm{E}}$ and $\delta_{\textrm{A}_1}$ versus $\delta_{\textrm{E}}$. In each case majority of the defects lie above the dotted line with a slope of unity, showing that $A_1$-like contributions dominate the $E$-like contributions. Applying a constant offset on the $V_{\textrm{A}_1}$ values could, in principle, shift additional points below the line but at the same time several of the points with low values of $V_{\textrm{A}_1}$ would also shift above the line. Testing for values of constant $A_1$-like strain offset between -30 and +30 GHz we find no constant offset value for which the majority of points lie below the line. Thus, the static strain is dominated by $A_1$-like perturbations. $\delta_{\textrm{A}_1}$ does not suffer from a potential constant offset, and so the presence of the majority of points above the line indicates that spectral diffusion is dominated by $A_1$-like perturbations.

\end{appendix}

\clearpage

\bibliography{bibliography_main}

\end{document}